\documentclass[11pt]{article}
\usepackage{amssymb}
\usepackage{amsfonts}
\usepackage{indentfirst}
\usepackage{amsopn}
\usepackage{amsmath}
\usepackage{amsthm}
\usepackage{amscd}
\usepackage{graphicx}
\usepackage{cite}

\begin{document}

\title{\bf Rogue periodic waves of the focusing NLS equation}

\author{Jinbing Chen$^{1}$ and Dmitry E. Pelinovsky$^{2}$ \\
{\small \it $^1$ School of Mathematics, Southeast University, Nanjing, Jiangsu 210096, P.R. China} \\
{\small \it $^2$ Department of Mathematics, McMaster University, Hamilton, Ontario, Canada, L8S 4K1 }  }

\date{\today}
\maketitle

\begin{abstract}
Rogue waves on the periodic background are considered for the nonlinear Schr\"{o}dinger
(NLS) equation in the focusing case. The two periodic wave solutions are
expressed by the Jacobian elliptic functions {\em dn} and {\em cn}. Both periodic waves
are modulationally unstable with respect to long-wave perturbations. Exact solutions
for the rogue waves on the periodic background are constructed by using the explicit expressions
for the periodic eigenfunctions of the Zakharov--Shabat spectral problem and
the Darboux transformations. These exact solutions labeled as {\em rogue periodic waves}
generalize the classical rogue wave (the so-called Peregrine's breather).
The magnification factor of the rogue periodic waves is computed as a function
of the wave amplitude (the elliptic modulus). Rogue periodic waves
constructed here are compared with the rogue wave patterns obtained numerically in recent
publications.
\end{abstract}

\section{Introduction}

Nonlinear waves in fluids are modeled by the nonlinear Schr\"{o}dinger (NLS) equation
in many physical situations. The same model is also relevant
to describe frequent occurrence of gigantic waves on the ocean's surface called
{\em rogue waves} \cite{Charif}. By its definition, a rogue wave
appears from nowhere and disappears without a trace \cite{Taki}. From a physical perspective,
the rogue waves emerge on the background of modulationally unstable nonlinear waves,
e.g. constant wave background, periodic waves, or quasi-periodic spatially-temporal patterns 
\cite{AZ1,AZ2,Calini}.

In what follows, we take the focusing NLS equation in the normalized form:
\begin{equation}
i u_t + u_{xx} + 2 |u|^2 u =0.
\label{1}
\end{equation}
The NLS equation (\ref{1}) appears as a compatibility condition of the following Lax pair
of linear equations on $\varphi \in \mathbb{C}^2$:
\begin{equation}\label{2}
\varphi_x=U\varphi,\qquad \qquad U=\left(\begin{array}{cc}
\lambda&u\\
-\bar{u}&-\lambda\\
\end{array}
\right)
\end{equation}
and
\begin{equation}\label{3}
\varphi_t=V\varphi,\qquad V = i \left(\begin{array}{cc}
2\lambda^2+|u|^2&u_x+2\lambda u\\
\bar{u}_x-2\lambda\bar{u}&-2\lambda^2-|u|^2\\
\end{array}
\right),
\end{equation}
where $\bar{u}$ is the conjugate of $u$. The first equation (\ref{2}) is usually referred to as
the Zakharov--Shabat spectral problem with the spectral parameter $\lambda$, whereas
the second equation (\ref{3}) determines the time evolution of the eigenfunctions
of the Zakharov--Shabat spectral problem.

The {\em classical rogue wave} up to the translations in the $(x,t)$ plane is given by
the exact rational solution of the NLS equation (\ref{1}):
\begin{equation}
\label{rogue-basic}
u(x,t) = \left[ 1 - \frac{4 (1+4it)}{1 + 4 x^2 + 16 t^2} \right] e^{2it}.
\end{equation}
As $|t| + |x| \to \infty$, the rogue wave (\ref{rogue-basic}) approaches the constant wave
background $u_0(x,t) = e^{2it}$. At the origin $(x,t) = (0,0)$, the rogue wave reaches the
maximum at $|u(0,0)| = 3$, from which the {\em magnification factor} of the
constant wave background is defined to be $M_0 = 3$. The rogue wave (\ref{rogue-basic}) was derived by
Peregrine \cite{Peregrine} as an outcome of the modulational instability of the constant wave
background of the focusing NLS equation (\ref{1}) and is sometimes referred to as {\em Peregrine's breather}.
More complicated rational solutions for rogue waves in the NLS equation (\ref{1}) were constructed by applications of the
multi-fold Darboux transformations \cite{Akh,Matveev,JYang}.

It is relatively less studied on how
to construct rogue waves on the non-constant background. Several recent publications offer different computational
tools in the context of rogue waves on the background of periodic or two-phase solutions.

Computations of rogue waves on the periodic background were performed in \cite{Kedziora},
where solutions of the Zakharov--Shabat spectral problem were computed numerically and
these approximations were substituted into the one-fold Darboux transformation. Since the spectral parameter was selected
at random in \cite{Kedziora} without connection to the band-gap spectrum of the Zakharov--Shabat spectral problem,
the resulting wave patterns do not represent accurately the rogue waves on the periodic background.

Computations of rogue waves on the two-phase background were achieved in \cite{CalSch} with a more accurate numerical scheme.
The authors constructed numerical solutions of the Zakharov--Shabat spectral problem
for particular branch points obtained also numerically, after which
the one-fold Darboux transformation was used.
The resulting wave patterns are periodic both in space and time with a rogue wave placed at the origin
and these patterns matched well with experimental data for rogue waves in fluids \cite{CalSch}.

General two-phase solutions of the focusing NLS equation (\ref{1}) were analyzed in \cite{Wright} and in \cite{Tovbis1,Tovbis2}.
Although the wave patterns for the general two-phase solutions are quasi-periodic both in space and time,
some parts of the quasi-periodic pattern look like rogue waves on the periodic background. In particular, 
the magnification factor of a rogue wave was computed as a ratio between the maximal amplitude and
the mean value of the two-phase solution \cite{Tovbis1,Tovbis2}.

Integrable turbulence and rogue waves were observed numerically in \cite{AZ2} during the modulational 
instability of the $dn$-periodic waves. In particular, the magnification factor of a rogue wave arising as a result 
of two-soliton collisions was observed to be two, in agreement with recent results of \cite{PelSl} 
obtained in the context of the focusing modified KdV equation. The rogue wave at the time of their maximal elevation was observed 
to have a quasi-rational profile similar to that of the Peregrine's breather \cite{AZ2}.

The purpose of our work is to obtain {\em exact analytical solutions} for the rogue waves on the periodic background,
which we name here as {\em rogue periodic waves}. We show how to compute exactly the branch points in the band-gap spectrum
of the Zakharov--Shabat problem associated with the periodic background, how to represent analytically the periodic and non-periodic
solutions of the Zakharov--Shabat problem, and how to generate accurately the rogue periodic waves by means of the one-fold
or two-fold Darboux transformations.

The standing periodic wave solutions to the focusing NLS equation (\ref{1}) can be represented in the form
\begin{equation}
\label{periodic-wave}
u(x,t) = U(x) e^{ict},
\end{equation}
where the periodic function $U$ satisfies the following second-order equation
\begin{equation}\label{19}
\frac{d^2 U}{d x^2} + 2 |U|^2 U = c  U.
\end{equation}
The second-order equation (\ref{19}) can be integrated to yield the following first-order invariant
\begin{equation}\label{19first}
\left| \frac{d U}{dx} \right|^2 + |U|^4 = c |U|^2 + d.
\end{equation}
Here $c$ and $d$ are real-valued constants. In addition to the standing periodic wave solutions (\ref{periodic-wave}),
there exist traveling periodic wave solutions with nontrivial dependence of the wave phase (see, e.g., in \cite{DS}).
For simplicity of our presentation, we only consider rogue waves on the standing periodic waves (\ref{periodic-wave}).

There are two particular families of the periodic wave
solutions in the focusing NLS equation (\ref{1})
expressed by the Jacobian elliptic functions $dn$ and $cn$ \cite{AS}.
The positive-definite $dn$-periodic waves are given by
\begin{equation}
\label{dn-periodic}
U(x) =  {\rm dn}(x;k), \quad
c = 2 - k^2, \quad d = -(1-k^2), \quad k \in (0,1),
\end{equation}
whereas the sign-indefinite $cn$-periodic waves are given by
\begin{equation}
\label{cn-periodic}
U(x) = k {\rm cn}(x;k), \quad c=2k^2-1, \quad d = k^2 (1-k^2), \quad k \in (0,1).
\end{equation}
In both cases, the periodic waves are even and centered at the point $x = 0$ thanks to the
translational invariance of the NLS equation (\ref{1}) in $x$. The parameter $k \in (0,1)$ is elliptic modulus
and in the limit $k \to 1$, both solutions converge to the normalized NLS soliton
\begin{equation}
\label{soliton}
U(x) = {\rm sech}(x), \quad c = 1, \quad d = 0.
\end{equation}
In the limit $k \to 0$, the $dn$ wave converges to the constant wave background
$u_0(x,t) = e^{2it}$, whereas the $cn$ wave converges to the zero background.

Spectral stability of the periodic waves in the focusing NLS equation was investigated
in details \cite{DS} (see also \cite{LeCoz,Lafortune}). It was found that
both $dn$- and $cn$-periodic waves are modulationally unstable with respect to the long-wave
perturbations (see review in \cite{BHJ}). The rogue periodic waves constructed in our work
are related to the modulational instability of the two periodic waves with respect to the
long-wave perturbations.

We will adopt the following definition of a rogue wave on the periodic background. For a given
periodic wave $u_{\rm per}(x,t) = U(x) e^{i c t}$, we say that the new solution $u$ is
a {\em rogue periodic wave} if it is different from an orbit of the periodic wave $u_{\rm per}$
due to translational and phase invariance of the NLS equation (\ref{1}) but
\begin{equation}
\label{rogue-wave-def}
\inf_{x_0,\alpha_0 \in \mathbb{R}} \sup_{x \in \mathbb{R}} \left| u(x,t) - U(x-x_0) e^{i \alpha_0} \right| \to 0
\quad \mbox{\rm as} \quad t \to \pm \infty.
\end{equation}
This definition corresponds to the common understanding of rogue waves as the waves that
appear from nowhere and disappear without a trace as the time evolves \cite{Taki}.

Our work relies on the analytical algorithm introduced recently in the context of periodic
waves in the focusing modified Korteweg--de Vries (KdV) equation \cite{ChenPel}.
First, by using the algebraic technique based on nonlinearization of the Lax pair \cite{Cao1},
we obtain the explicit expressions for the branch points $\lambda$ with ${\rm Re}(\lambda) > 0$
and the periodic eigenfunctions in the Zakharov--Shabat spectral problem (\ref{2})
associated with the $dn$- and $cn$-periodic waves. For each periodic eigenfunction,
we construct the second, linearly independent solution to the linear system (\ref{2})--(\ref{3}),
which is not periodic but linearly growing in $(x,t)$. Finally,
substituting non-periodic solutions to the linear system (\ref{2})--(\ref{3}) into the one-fold and two-fold
Darboux transformations \cite{GuHuZhou} yields the rogue periodic waves in the sense of the definition (\ref{rogue-wave-def}).

The paper is organized as follows. Section 2 reports construction of the
periodic eigenfunctions of the linear system (\ref{2})--(\ref{3}).
Section 3  describes construction of the rogue periodic waves.
Section 4 concludes the paper with further discussions.

\section{Periodic eigenfunctions of the Lax pair}

The algebraic technique based on the nonlinearization of the Lax pair was introduced in \cite{Cao1}.
It was implemented for the linear system (\ref{2})--(\ref{3}) in \cite{ZhouStudies,ZhouJMP}.
Here we use this algebraic technique for a novel purpose of constructing the explicit expressions
for periodic eigenfunctions of the Zakharov--Shabat spectral problem
associated with the periodic wave solutions (\ref{dn-periodic}) and (\ref{cn-periodic}).

\subsection{Nonlinearization of the Lax pair}

We introduce the following constraint \cite{ZhouStudies,ZhouJMP},
\begin{equation}\label{4}
u = p_1^2+\bar{q}_1^2,
\end{equation}
between the potential $u$ and a particular nonzero solution $\varphi = (p_1,q_1)^T$
of the linear system (\ref{2})--(\ref{3}) for $\lambda = \lambda_1$, where $\lambda_1 \in \mathbb{C}$ is fixed arbitrarily.

Substituting (\ref{4}) into the spectral problem (\ref{2}) yields a finite-dimensional Hamiltonian system
in complex variables
\begin{equation}\label{5}
\frac{d p_1}{d x} = \frac{\partial H}{\partial q_1}, \quad
\frac{d q_1}{d x} = - \frac{\partial H}{\partial p_1},
\end{equation}
which is associated with the real-valued Hamiltonian function
\begin{equation}\label{6}
H = \lambda_1p_1q_1 + \bar{\lambda}_1\bar{p}_1\bar{q}_1 + \frac12(p_1^2+\bar{q}_1^2)(\bar{p}_1^2+q_1^2).
\end{equation}
There exists two constants of motion for the system (\ref{5})--(\ref{6}):
\begin{equation}\label{10}
F_0=i(p_1q_1-\bar{p}_1\bar{q}_1),
\end{equation}
\begin{equation}\label{11}
F_1=\lambda_1p_1q_1+\bar{\lambda}_1\bar{p}_1\bar{q}_1+\frac12(|p_1|^2+|q_1|^2)^2,
\end{equation}
Indeed, $F_0$ is constant in $x$ due to the following cancelation:
\begin{eqnarray*}
\frac{d F_0}{dx} & = & iq_1 \left[ \lambda_1 p_1 + (p_1^2+\bar{q}_1^2) q_1 \right] + i p_1 \left[ -\lambda_1 q_1 - (\bar{p}_1^2+q_1^2) p_1 \right] \\
& \phantom{t} & -i\bar{q}_1 \left[ \bar{\lambda}_1 \bar{p}_1 + (\bar{p}_1^2+q_1^2) \bar{q}_1 \right] -
i \bar{p}_1 \left[ -\bar{\lambda}_1 \bar{q}_1 - (p_1^2+\bar{q}_1^2) \bar{p}_1 \right] = 0,
\end{eqnarray*}
whereas $F_1$ is constant in $x$ because it is related to the constant values of $H$ and $F_0$ as follows:
\begin{eqnarray}
H - F_1 =
\frac{1}{2} \left[ |p_1|^4 + p_1^2 q_1^2 + \bar{p}_1^2 \bar{q}_1^2 + |q_1|^4 \right] - \frac{1}{2} \left[ |p_1|^4 + 2 |p_1|^2 |q_1|^2 + |q_1|^4 \right] =
-\frac{1}{2} F_0^2.\label{relation-H-F}
\end{eqnarray}

Substituting (\ref{4}) into the time-evolution system (\ref{3}) yields another Hamiltonian system
\begin{equation}\label{a1}
\frac{d p_1}{dt} = \frac{\partial K}{\partial q_1}, \quad
\frac{d q_1}{dt} = -\frac{\partial K}{\partial p_1},
\end{equation}
associated with the real-valued Hamiltonian function
\begin{eqnarray}
\nonumber
K & = & i \left[
2\lambda_1^2p_1q_1-2\bar{\lambda}_1^2\bar{p}_1\bar{q}_1+|p_1^2+\bar{q}_1^2|^2(p_1q_1-\bar{p}_1\bar{q}_1) \right. \\
\label{a2}
& \phantom{t} & \left.
+ (\lambda_1p_1^2-\bar{\lambda}_1\bar{q}_1^2)(\bar{p}_1^2+q_1^2)+(p_1^2+\bar{q}_1^2)(\lambda_1q_1^2-\bar{\lambda}_1\bar{p}_1^2) \right].
\end{eqnarray}
In the derivation of system (\ref{a1})--(\ref{a2}) from system (\ref{3}), we have used the constraint (\ref{4})
and the following constraint
\begin{eqnarray}
\nonumber
u_x & = & 2 p_1 (\lambda_1 p_1 + (p_1^2 + \bar{q}_1^2) q_1) - 2 \bar{q}_1 (\bar{\lambda}_1 \bar{q}_1 + (p_1^2 + \bar{q}_1^2) \bar{p}_1) \\
\label{4a}
& = & 2 (\lambda_1 p_1^2 - \bar{\lambda}_1 \bar{q}_1^2) + 2 (p_1^2 + \bar{q}_1^2) (p_1 q_1 - \bar{p}_1 \bar{q}_1),
\end{eqnarray}
which follows from the differentiation of (\ref{4}) and the substitution of (\ref{5})--(\ref{6}).

The two quantities $F_0$ and $F_1$ given by (\ref{10}) and (\ref{11}) are
constants of motion for the system (\ref{a1})--(\ref{a2}). Indeed, $F_0$ is constant in $t$
due to the following cancelation:
\begin{eqnarray*}
\frac{d F_0}{dt} & = & - q_1 \left[ (2\lambda_1^2 + |u|^2) p_1 + (u_x + 2 \lambda_1 u) q_1 \right]
 - p_1 \left[ -(2\lambda_1^2+|u|^2) q_1 + (\bar{u}_x - 2 \lambda_1 \bar{u}) p_1 \right] \\
& \phantom{t} & - \bar{q}_1 \left[ (2\bar{\lambda}_1^2 + |u|^2) \bar{p}_1 + (\bar{u}_x + 2 \bar{\lambda}_1 \bar{u}) \bar{q}_1 \right]
- \bar{p}_1 \left[ -(2\bar{\lambda}_1^2+|u|^2) \bar{q}_1 + (u_x - 2 \bar{\lambda}_1 u) \bar{p}_1 \right] \\
& = & - \left[ \bar{u} u_x + u \bar{u}_x + 2 u (\lambda_1 q_1^2 - \bar{\lambda}_1 \bar{p}_1^2) + 2 \bar{u} (\bar{\lambda}_1 \bar{q}_1^2 - \lambda_1^2 p_1^2) \right] \\ &=& 0,
\end{eqnarray*}
where the last identity follows by (\ref{4a}) and its complex conjugate. In order to prove that
$F_1$ is constant in $t$, it is sufficient to prove
that $H$ is constant in $t$, thanks to the relation (\ref{relation-H-F}) between $H$, $F_0$, and $F_1$.
To do so, we introduce the complex Poisson bracket in $\mathbb{C}^2$ associated with the
symplectic structures of the systems (\ref{5})--(\ref{6}) and (\ref{a1})--(\ref{a2}):
$$
\{ f, g \} := \frac{\partial f}{\partial p_1}\frac{\partial g}{\partial q_1}-\frac{\partial f}{\partial q_1}\frac{\partial g}{\partial p_1}
+\frac{\partial f}{\partial \bar{p}_1}\frac{\partial g}{\partial \bar{q}_1}-\frac{\partial f}{\partial \bar{q}_1}\frac{\partial g}{\partial \bar{p}_1}.
$$
Then, it follows from (\ref{6}), (\ref{a2}), and (\ref{4a}) that
\begin{eqnarray*}
\{H,K\} &=& i \left[ (\lambda_1q_1 + \bar{u}p_1)((2\lambda_1^2+|u|^2)p_1+(u_x+2\lambda_1u)q_1) \right. \\&&
\left. +(\lambda_1p_1+uq_1)((\bar{u}_x-2\lambda_1\bar{u})p_1-(2\lambda_1^2+|u|^2)q_1) \right. \\&&
\left. -(\bar{\lambda}_1\bar{q}_1 + u\bar{p}_1)((2\bar{\lambda}_1^2+|u|^2)\bar{p}_1+(\bar{u}_x+2\bar{\lambda}_1\bar{u})\bar{q}_1) \right. \\&&
\left. -(\bar{\lambda}_1\bar{p}_1+\bar{u}\bar{q}_1)((u_x-2\bar{\lambda}_1u)\bar{p}_1-(2\bar{\lambda}_1^2+|u|^2)\bar{q}_1) \right]\\
&=& i \left[ \lambda_1(u_x q_1^2+\bar{u}_x p_1^2)-\bar{\lambda}_1(\bar{u}_x \bar{q}_1^2 + u_x \bar{p}_1^2)
+(\bar{u}u_x + \bar{u}_x u) (p_1 q_1 - \bar{p}_1 \bar{q}_1) \right] \\
&=& i \left[ u_x (\lambda_1 q_1^2 - \bar{\lambda}_1 \bar{p}_1^2 + \bar{u}(p_1 q_1 - \bar{p}_1 \bar{q}_1)) +
\bar{u}_x (\lambda_1 p_1^2 - \bar{\lambda}_1 \bar{q}_1^2 + u (p_1 q_1 - \bar{p}_1 \bar{q}_1)) \right] \\
&=& 0.
\end{eqnarray*}
Since $H$ and $K$ commutes, then $H$ is constant in $t$ and $K$ is constant in $x$.

Let us summarize this first step of our computational algorithm. We have obtained
two commuting Hamiltonian systems (\ref{5})--(\ref{6}) and (\ref{a1})--(\ref{a2})
on the eigenfunction $(p_1,q_1)$ of the linear system (\ref{2})--(\ref{3}) associated
with the eigenvalue $\lambda_1$ and the potential $u$ related to $(p_1,q_1)$ by
the algebraic constraint (\ref{4}). In the next step, we obtain differential constraints
on the potential $u$ from the integrability scheme for the Hamiltonian system (\ref{5})--(\ref{6}).
One differential constraint is already obtained in (\ref{4a}), which can be written in the equivalent form:
\begin{eqnarray}
\label{4a-new}
\frac{du}{dx}  = 2 (\bar{\lambda}_1 p_1^2 - \lambda_1 \bar{q}_1^2)
+ 2 (p_1^2 + \bar{q}_1^2) (\lambda_1 - \bar{\lambda}_1 + p_1 q_1 - \bar{p}_1 \bar{q}_1),
\end{eqnarray}
where the ordinary derivatives are used for convenience and the time dependence is also assumed.
We will obtain other differential constraints on $u$, which resemble the second-order
equation (\ref{19}) and its first-order invariant (\ref{19first}) for the periodic waves
(\ref{dn-periodic}) and (\ref{cn-periodic}). From here, we will conclude that the
differential constraints are satisfied if $u$ is the periodic wave of the NLS equation (\ref{1})
given by (\ref{periodic-wave}). Since $u$ is a compatibility condition of the linear system (\ref{2})--(\ref{3}),
we do not have to deal with the commuting Hamiltonian
system (\ref{a1})--(\ref{a2}), as the time evolution of $(p_1,q_1)$ can be deduced
from the algebraic constraint (\ref{4}) and the conserved quantities $F_0$ and $F_1$ in (\ref{10}) and (\ref{11}).

We note here that the extension of the constraint (\ref{4}) is possible with several solutions
of the linear system (\ref{2})--(\ref{3}) for distinct values of $\lambda$ \cite{ZhouJMP,ZhouStudies}.
This multi-function construction is related to the multi-phase (quasi-periodic) solutions of the NLS equation (\ref{1})
expressed by the Riemann's Theta function \cite{MatveevBook}. It remains open due to higher computational
difficulties to obtain rogue waves on the background of multi-phase solutions.

\subsection{Differential constraints on the potential $u$}

Hamiltonian system (\ref{5})--(\ref{6}) is a compatibility condition for the Lax equation
\begin{equation}\label{8}
\frac{d}{dx} W(\lambda) = [Q(\lambda),W(\lambda)], \quad \lambda \in \mathbb{C},
\end{equation}
where
\begin{equation}\label{7}
Q(\lambda) = \left(\begin{array}{cc}
\lambda & p_1^2+\bar{q}_1^2\\
-\bar{p}_1^2-q_1^2 & -\lambda \end{array} \right), \quad
W(\lambda) =\left(\begin{array}{cc}
W_{11}(\lambda) & W_{12}(\lambda) \\
\overline{W_{12}(-\bar{\lambda})} & -\overline{W_{11}(-\bar{\lambda})} \end{array}\right)
\end{equation}
with
\begin{eqnarray*}
W_{11}(\lambda) = 1-\frac{p_1q_1}{\lambda-\lambda_1} + \frac{\bar{p}_1\bar{q}_1}{\lambda+\bar{\lambda}_1},
\end{eqnarray*}
and
\begin{eqnarray*}
W_{12}(\lambda) = \frac{p_1^2}{\lambda-\lambda_1} + \frac{\bar{q}_1^2}{\lambda+\bar{\lambda}_1}.
\end{eqnarray*}
In particular, the $(1,2)$-entry of the Lax equation (\ref{8}) is rewritten in the form:
\begin{equation}
\label{Dub}
\frac{d}{dx} W_{12}(\lambda) = 2 \lambda W_{12}(\lambda) - 2(p_1^2+\bar{q}_1^2) W_{11}(\lambda).
\end{equation}

We rewrite $W_{11}(\lambda)$ and $W_{12}(\lambda)$ in terms of $u$ and constants of motion
$F_0$ and $F_1$ by using relations (\ref{4}), (\ref{10}), (\ref{11}), and (\ref{4a-new}).
Some routine computations yields the following explicit expressions:
\begin{eqnarray*}
W_{11}(\lambda) & = & 1 - \frac{\lambda (p_1 q_1 - \bar{p}_1 \bar{q}_1) + \bar{\lambda}_1 p_1 q_1 + \lambda_1 \bar{p}_1\bar{q}_1}{
(\lambda - \lambda_1) (\lambda + \bar{\lambda}_1)} \\
& = & 1 + \frac{i F_0 (\lambda - \lambda_1 + \bar{\lambda}_1) + \frac{1}{2} F_0^2 - F_1 + \frac{1}{2} |u|^2}{
(\lambda - \lambda_1) (\lambda + \bar{\lambda}_1)}
\end{eqnarray*}
and
\begin{eqnarray*}
W_{12}(\lambda) & = & \frac{\lambda (p_1^2 + \bar{q}_1^2) + \bar{\lambda}_1 p_1^2 - \lambda_1 \bar{q}_1^2}{(\lambda - \lambda_1) (\lambda + \bar{\lambda}_1)}\\
& = & \frac{(\lambda - \lambda_1 + \bar{\lambda}_1 + i F_0) u + \frac{1}{2} u_x}{(\lambda - \lambda_1) (\lambda + \bar{\lambda}_1)}.
\end{eqnarray*}
Substituting these expressions for $W_{11}(\lambda)$ and $W_{12}(\lambda)$
into equation (\ref{Dub}) yields the following differential constraint on $u$:
\begin{equation}
\label{connect-1}
\frac{d^2 u}{d x^2} + 2i (F_0 + i \lambda_1 - i \bar{\lambda}_1) \frac{du}{dx} + 2 |u|^2 u =
4 \left( |\lambda_1|^2 + F_1 - \frac{1}{2} F_0^2
+ i F_0 (\lambda_1 - \bar{\lambda}_1) \right) u.
\end{equation}
This equation is to be compared with the second-order differential equation (\ref{19}).
In order to obtain the first-order invariant (\ref{19first}), we consider the determinant of $W(\lambda)$.
As is well-known \cite{Tu},
\begin{itemize}
\item $\det W(\lambda)$ has simple poles at $\lambda = \lambda_1$ and $\lambda = -\bar{\lambda}_1$
\item $\det W(\lambda)$ is  independent of $x$ and $t$ as it is related to the integrals of motion $F_0$ and $F_1$
for the Hamiltonian systems (\ref{5})--(\ref{6}) and (\ref{a1})--(\ref{a2}).
\end{itemize}
These two properties are verified with the following explicit computation:
\begin{eqnarray*}
\det W(\lambda) & = & - W_{11}(\lambda) \overline{W_{11}(-\bar{\lambda})} - W_{12}(\lambda) \overline{W_{12}(-\bar{\lambda})} \\
& = & -1 + \frac{2p_1q_1}{\lambda-\lambda_1}-\frac{2\bar{p}_1\bar{q}_1}{\lambda+\bar{\lambda}_1}+
\frac{(|p_1|^2+|q_1|^2)^2}{(\lambda-\lambda_1)(\lambda+\bar{\lambda}_1)} \\
& = & -\frac{(\lambda-\lambda_1)(\lambda + \bar{\lambda}_1) + 2 i (\lambda - \lambda_1 + \bar{\lambda}_1) F_0 - 2F_1}{
(\lambda - \lambda_1) (\lambda + \bar{\lambda}_1)}.
\end{eqnarray*}
On the other hand, since $\overline{W_{11}(-\bar{\lambda})} = W_{11}(\lambda)$, we can also use the explicit expressions
for $W_{11}(\lambda)$ and $W_{12}(\lambda)$ and rewrite $\det W(\lambda)$ in the following form:
\begin{eqnarray*}
\det W(\lambda) & = & - \left[ 1 + \frac{i F_0 (\lambda - \lambda_1 + \bar{\lambda}_1) + \frac{1}{2} F_0^2 - F_1 + \frac{1}{2} |u|^2}{
(\lambda - \lambda_1) (\lambda + \bar{\lambda}_1)} \right]^2 \\
& \phantom{t} &
+ \frac{\left[ (\lambda - \lambda_1 + \bar{\lambda}_1 + i F_0) u + \frac{1}{2} u_x\right]
\left[ (\lambda - \lambda_1 + \bar{\lambda}_1 + i F_0) \bar{u} - \frac{1}{2} \bar{u}_x\right]}{
(\lambda - \lambda_1)^2 (\lambda + \bar{\lambda}_1)^2}.
\end{eqnarray*}
The representation above has double poles at $\lambda = \lambda_1$ or $\lambda = -\bar{\lambda}_1$,
which are identically zero due to the properties of $\det W(\lambda)$. Removing the double poles
at $\lambda = \lambda_1$ or $\lambda = -\bar{\lambda}_1$ yields the following two
differential constraints on $u$:
\begin{eqnarray*}
&& \left| \frac{du}{dx} \right|^2  + |u|^4 - 2 (i F_0 + \bar{\lambda}_1) (\bar{u} u_x - \bar{u}_x u) \\
&& + 2 \left( 3 F_0^2 - 2 F_1 - 2 i F_0 \bar{\lambda}_1 - 2 \bar{\lambda}_1^2 \right) |u|^2 +
\left( F_0^2 + 2 i F_0 \bar{\lambda}_1 - 2 F_1 \right)^2 = 0
\end{eqnarray*}
and
\begin{eqnarray*}
&& \left| \frac{du}{dx} \right|^2  + |u|^4 - 2 (i F_0 - \lambda_1) (\bar{u} u_x - \bar{u}_x u)\\
&& + 2 \left( 3 F_0^2 - 2 F_1 + 2 i F_0 \lambda_1 - 2 \lambda_1^2 \right) |u|^2 +
\left( F_0^2 - 2 i F_0 \lambda_1 - 2 F_1 \right)^2 = 0.
\end{eqnarray*}

Let us represent $\lambda_1=\alpha+i\beta$ with $\alpha,\beta \in \mathbb{R}$.
Subtracting one differential constraint from the other one yields the following simpler
constraint on $u$:
\begin{equation}
\bar{u} \frac{d u}{dx} - u \frac{d \bar{u}}{dx} = 2 i (2 \beta - F_0) |u|^2 + 2 i F_0 (F_0^2 + 2 F_0 \beta - 2 F_1).
\label{connect-2}
\end{equation}
Substituting this constraint in either of the two differential constraints above
yields another equivalent differential constraint on $u$:
\begin{eqnarray}
\nonumber
&& \left| \frac{du}{dx} \right|^2 + |u|^4 + 2 \left( F_0^2 - 2 F_1 + 4  F_0 \beta  - 2 \alpha^2 - 2 \beta^2 \right) |u|^2 \\
&& + \left( F_0^2 + 2 F_0 \beta - 2 F_1 \right) \left( 5 F_0^2 - 2 F_0 \beta - 2 F_1 \right)
- 4 F_0^2 \alpha^2 = 0.
\label{connect-3}
\end{eqnarray}
The latter equation is to be compared with the first-order invariant (\ref{19first}).

We note here that the differential constraints (\ref{connect-1}), (\ref{connect-2}), and (\ref{connect-3})
are more general than the differential equations (\ref{19}) and (\ref{19first}).
In particular, the constraints can be used to address traveling periodic wave
solutions of the NLS equation (\ref{1}) with a nontrivial dependence of the wave phase \cite{DS}.
The corresponding extension is straightforward and is omitted for the sake of clarity.

\subsection{$dn$- and $cn$-periodic waves}

Let us connect the differential equations (\ref{19}) and (\ref{19first})
for the periodic waves (\ref{dn-periodic}) and (\ref{cn-periodic}) with
the differential constraints (\ref{connect-1}), (\ref{connect-2}), and (\ref{connect-3}).
Both periodic waves give zero in the left-hand side of equation (\ref{connect-2}).
Hence, we obtain the following relations:
\begin{equation}
\label{relation-F0-F1}
F_0 = 2 \beta, \quad \beta (F_1 - 4 \beta^2) = 0.
\end{equation}
Comparing coefficients in (\ref{19}) and (\ref{19first}) with the coefficients in (\ref{connect-1})
 and (\ref{connect-3}) yields
\begin{equation}
\label{relation-c-d}
c = 4 (\alpha^2 - 5 \beta^2 + F_1 ), \quad d = 4 (4 \alpha^2 \beta^2 - (F_1 - 4 \beta^2)(F_1 - 8 \beta^2) ),
\end{equation}
where the relations (\ref{relation-F0-F1}) have been taken into account.
The second equation in (\ref{relation-F0-F1}) can be satisfied with two choices:
either $\beta = 0$ or $\beta \neq 0$ and $F_1 = 4 \beta^2$. Both choices
are relevant for the periodic waves (\ref{dn-periodic}) and (\ref{cn-periodic}).

If $\beta = 0$, then relations (\ref{relation-F0-F1}) yield $F_0 = 0$, whereas
relations (\ref{relation-c-d}) yield
\begin{equation}\label{20}
c=4(\alpha^2+F_1),\qquad d=-4F_1^2.
\end{equation}
Since $d < 0$, we can only compare these expressions for $(c,d)$ with those
for the $dn$-periodic wave in (\ref{dn-periodic}). This yields the following expressions
for $F_1$ and $\lambda_1 = \alpha$ in terms of the elliptic modulus $k \in (0,1)$:
$$
F_1 = \pm \frac{1}{2} \sqrt{1-k^2}, \quad \lambda_1^2 = \frac{1}{4} \left[ 2 - k^2 \mp 2 \sqrt{1-k^2} \right].
$$
The expressions for $\lambda_1$ give two real eigenvalues in the right half-plane
\begin{equation}\label{lambda-1-dn}
\lambda_{\pm} := \frac{1}{2} (1 \pm \sqrt{1-k^2})
\end{equation}
and two symmetric eigenvalues $-\lambda_{\pm}$ in the left half-plane.

If $\beta \neq 0$, then relations (\ref{relation-F0-F1}) yield
$F_0 = 2\beta$ and $F_1 = 4\beta^2$, whereas relations (\ref{relation-c-d}) yield
\begin{equation}\label{21}
c=4(\alpha^2-\beta^2), \quad d=16\alpha^2\beta^2.
\end{equation}
Since $d > 0$, we can only compare these expressions for $(c,d)$ with those
for the $cn$-periodic wave in (\ref{cn-periodic}). This yields the following expression
for $\lambda_1 = \alpha + i \beta$ in terms of the elliptic modulus $k$:
$$
\lambda_1^2 = \frac{1}{4} \left[ 2 k^2 - 1 \pm 2 i k \sqrt{1-k^2} \right].
$$
The expressions for $\lambda_1$ give the eigenvalue in the first quadrant
\begin{equation}\label{lambda-1-cn}
\lambda_I := \frac{1}{2} \left[ k + i \sqrt{1-k^2} \right].
\end{equation}
and three symmetric eigenvalues $\bar{\lambda}_I$, $-\lambda_I$, and $-\bar{\lambda}_I$
in the other three quadrants. 

\subsection{Periodic eigenfunctions}

We complete the last step of the algorithm and obtain identities for the periodic eigenfunctions
of the Zakharov--Shabat spectral problem (\ref{2}) associated with the periodic wave $u$.
These identities arise due to the constraints imposed on the periodic wave $u$ and the eigenfunction
$(p_1,q_1)$. In particular, relations (\ref{4}), (\ref{10}), and (\ref{4a}) set up the
following linear system for $(p_1^2,\bar{q}_1^2)$:
$$
\left\{ \begin{array}{l} p_1^2 + \bar{q}_1^2 = u, \\
\lambda_1 p_1^2 - \bar{\lambda}_1 \bar{q}_1^2 = \frac{1}{2} u_x + i u F_0 \end{array} \right.
$$
Since $\lambda_1 = \alpha + i \beta$ and $F_0 = 2 \beta$, we can obtain
the squared eigenfunctions explicitly as follows:
\begin{equation}\label{24}
p_1^2=\frac{2\lambda_1u+u_x}{2(\lambda_1+\bar{\lambda}_1)},\qquad
\bar{q}_1^2=\frac{2\bar{\lambda}_1 u - u_x}{2(\lambda_1+\bar{\lambda}_1)},
\end{equation}
In what follows, it will be useful to separate the time dependence
from the periodic wave $u(x,t) = U(x) e^{i ct}$. Then, representation
(\ref{24}) implies the following time dependence of the periodic eigenfunction $(p_1,q_1)$:
\begin{equation}
\label{periodic-wave-t}
p_1(x,t) = P_1(x) e^{i c t/2}, \quad q_1(x,t) = Q_1(x) e^{-i c t/2}.
\end{equation}
Since $U$ is real, the squared complex eigenfunctions are expressed by
\begin{equation}\label{24-t}
P_1(x)^2 = \frac{2\lambda_1 U(x) + U'(x)}{2(\lambda_1+\bar{\lambda}_1)},\qquad
\bar{Q}_1(x)^2 = \frac{2\bar{\lambda}_1 U(x) - U'(x)}{2(\lambda_1+\bar{\lambda}_1)}.
\end{equation}

For the $dn$-periodic waves (\ref{dn-periodic}), we have $U(x) = {\rm dn}(x;k)$,
$\beta = 0$, and $\alpha = \lambda_+$ given by (\ref{lambda-1-dn}). Since
$F_0 = 0$ and $H = F_1 = -\frac{1}{2} \sqrt{1-k^2}$ in this case,
the representations (\ref{6}) and (\ref{10}) yield
$4 \lambda_+ p_1 q_1 = 4 \lambda_+ \bar{p}_1 \bar{q}_1 = -|u|^2 - \sqrt{1-k^2}$,
whereas the representation (\ref{11}) yields
$(|p_1|^2 + |q_1|^2)^2 = |u|^2$. The previous two relations can be rewritten explicitly as
\begin{equation}\label{a8-t}
P_1(x) Q_1(x) = -\frac{1}{4 \lambda_+} \left[ U(x)^2 + \sqrt{1-k^2} \right]
\end{equation}
and
\begin{equation} \label{a6}
P_1(x)^2 + Q_1(x)^2 = U(x).
\end{equation}
It follows from (\ref{24-t}) with $\lambda_1 = \lambda_+$
that the squared eigenfunctions $P_1^2$ and $Q_1^2$ are real. Then it follows from
(\ref{a6}) with $U(x) = {\rm dn}(x;k) > 0$ that $P_1$ and $Q_1$ are real.

For the $cn$-periodic waves (\ref{cn-periodic}), we have $U(x) = k {\rm cn}(x;k)$,
$\alpha = \frac{1}{2} k$ and $\beta = \frac{1}{2} \sqrt{1-k^2}$,
so that $\lambda_I = \alpha + i \beta$ is
given by (\ref{lambda-1-cn}). Since $F_0 = 2 \beta$, $F_1 = 4 \beta^2$, and $H = 2 \beta^2$,
it follows from (\ref{6}) and (\ref{10}) that
${\rm Re}(p_1q_1) = -\frac{1}{2k} |u|^2$ and ${\rm Im}(p_1q_1) = -\frac{1}{2} \sqrt{1-k^2}$
so that $2k p_1 q_1 = -|u|^2 - i k \sqrt{1 - k^2}$, which can be written explicitly as
\begin{equation}
\label{28-t}
P_1(x) Q_1(x) = -\frac{1}{2k} \left[ U(x)^2 + i k \sqrt{1 - k^2} \right].
\end{equation}
On the other hand, it follows from (\ref{11}) and (\ref{28-t}) that
$(|p_1|^2 + |q_1|^2)^2 = 1 - k^2 + |u|^2$, hence
\begin{equation} \label{29}
|P_1(x)|^2 + |Q_1(x)|^2={\rm dn}(x;k).
\end{equation}
Furthermore, by using $F_1 = F_0^2$, we derive another relation
\begin{equation*}
\lambda_1p_1q_1+\bar{\lambda}_1\bar{p}_1\bar{q}_1+p_1^2q_1^2+\bar{p}_1^2\bar{q}_1^2+\frac12(|p_1|^2-|q_1|^2)^2=0,
\end{equation*}
which yields $(|p_1|^2-|q_1|^2)^2 = |u|^2 - |u|^4/k^2$ due to (\ref{lambda-1-cn}) and (\ref{28-t}).
Taking the negative square root yields the relation
\begin{equation}\label{47}
|P_1(x)|^2 - |Q_1(x)|^2 = - k {\rm sn}(x;k) {\rm cn}(x;k).
\end{equation}
The reason why the negative square root must be taken is explained from the following argument.
By using (\ref{24-t}), we know that
\begin{equation*}
|P_1(x)|^4 = \frac{1}{4k^2} \left[ (kU(x) + U'(x))^2 + (1-k^2) U(x)^2 \right]
\end{equation*}
and
\begin{equation*}
|Q_1(x)|^4 =  \frac{1}{4k^2} \left[ (kU(x) - U'(x))^2 + (1-k^2) U(x)^2 \right],
\end{equation*}
where $U'(x) = -k {\rm sn}(x;k) {\rm dn}(x;k)$. Since ${\rm dn}(x;k)>0$,
we have $|P(x)| < |Q(x)|$ if ${\rm sn}(x;k) {\rm cn}(x;k) > 0$. This is true for the negative square root
in (\ref{47}) and false for the positive square root.

It follows from  (\ref{29}) and (\ref{47}) that
\begin{equation}\label{d3}
|P_1(x)|^2=\frac{{\rm dn}(x;k)-k{\rm sn}(x;k){\rm cn}(x;k)}{2},\quad
|Q_1(x)|^2=\frac{{\rm dn}(x;k)+k{\rm sn}(x;k){\rm cn}(x;k)}{2}.
\end{equation}
Furthermore, it follows from (\ref{lambda-1-cn}) and (\ref{24-t}) that
\begin{eqnarray*}
P_1(x)^2 \bar{Q}_1(x)^2 = \frac14 \left[{\rm cn}^2(x;k)-{\rm sn}^2(x;k) {\rm dn}^2(x;k) +
 2 i\sqrt{1-k^2}{\rm sn}(x;k){\rm cn}(x;k){\rm dn}(x;k) \right]
\end{eqnarray*}
Taking the negative square root yields the following relation:
\begin{equation}\label{34}
P_1(x) \bar{Q}_1(x) =-\frac12 {\rm cn}(x;k){\rm dn}(x;k) -\frac{i}{2} \sqrt{1-k^2} {\rm sn}(x;k).
\end{equation}
The choice of the negative square root is explained as follows.
Combining (\ref{28-t}) with (\ref{34}) yields
\begin{eqnarray*}
P_1(x)^2 |Q_1(x)|^2 & = & \frac14 (k{\rm cn}^3(x;k){\rm dn}(x;k)-(1-k^2){\rm sn}(x;k)) \\
& \phantom{t} & + \frac{i}{4} \sqrt{1-k^2} {\rm cn}(x;k) ({\rm dn}(x;k)+k{\rm sn}(x;k){\rm cn}(x;k)),
\end{eqnarray*}
which coincides with the expression for $P_1(x)^2 |Q_1(x)|^2$ obtained
from (\ref{24-t}) and (\ref{d3}). In the case of the positive
square root in (\ref{34}), the expression for $P_1(x)^2 |Q_1(x)|^2$
obtained from (\ref{28-t}) would be negative to the one
obtained from (\ref{24-t}). Thus, the negative sign in
(\ref{34}) is justified.

\section{Construction of rogue periodic waves}

The rogue periodic waves can be constructed with the one-fold or two-fold Darboux transformations involving
the periodic eigenfunction $(p_1,q_1)$ for the eigenvalue $\lambda_1$ and possibly
another periodic eigenfunction $(p_2,q_2)$ for the eigenvalue $\lambda_2$,
since two eigenvalues with positive real parts were identified for each periodic wave.
However, such Darboux transformations recover only trivial solutions produced
from the periodic wave by means of spatial translations. In order to obtain nontrivial solutions
which corresponds to a rogue wave on the periodic background in the sense of the definition (\ref{rogue-wave-def}),
we will obtain the non-periodic solutions
to the linear system (\ref{2})--(\ref{3}) for the same eigenvalue $\lambda_1$.

\subsection{Non-periodic solutions of the Lax pair}

Let $u$ be a periodic wave of the NLS equation (\ref{1}) and $(p_1,q_1)$ be the $x$-periodic eigenfunctions
of the linear system (\ref{2}) and (\ref{3}) with $\lambda = \lambda_1$. Let us now construct the
second, linearly independent solution of the  linear system (\ref{2})--(\ref{3}) with $\lambda = \lambda_1$
denoted by $(\textsf{p}_1,\textsf{q}_1)$.
If $\lambda_1$ is a simple eigenvalue of the periodic spectral problem (\ref{2}), then $(\textsf{p}_1,\textsf{q}_1)$
is not periodic in $x$. We set
\begin{equation}\label{49}
\textsf{p}_1 = \frac{\theta - 1}{q_1}, \qquad \textsf{q}_1 = \frac{\theta + 1}{p_1},
\end{equation}
so that the Wronskian between the two linearly independent solutions $(p_1,q_1)$ and $(\textsf{p}_1,\textsf{q}_1)$
is normalized by $2$. Substituting (\ref{49}) into (\ref{2}) yields a first-order equation on $\theta$:
\begin{equation}\label{50}
\frac{d \theta}{dx} = \theta \frac{uq_1^2-\bar{u}p_1^2}{p_1q_1}+\frac{uq_1^2+\bar{u}p_1^2}{p_1q_1}.
\end{equation}
Note that this differential equation is invariant with respect to $t$
thanks to the representation (\ref{periodic-wave}) and (\ref{periodic-wave-t}).
Hence we write
\begin{equation}\label{50-t}
\frac{d \theta}{dx} = \theta U \frac{Q_1^2 - P_1^2}{P_1 Q_1} + U \frac{Q_1^2 + P_1^2}{P_1 Q_1},
\end{equation}
where $U$ is real for both periodic waves (\ref{dn-periodic}) and (\ref{cn-periodic}).

For the $dn$-periodic waves with $U(x) = {\rm dn}(x;k)$, it follows from (\ref{lambda-1-dn}) and (\ref{24-t}) that
\begin{equation}
\label{g1}
4 \lambda_+ \left[ P_1(x)^2 - Q_1(x)^2 \right] = 2 U'(x),
\end{equation}
where $\lambda_1 = \lambda_+$ is used. Together with (\ref{a8-t}) and (\ref{a6}) for real $P_1$ and $Q_1$,
we rewrite the differential relation (\ref{50-t}) in the explicit form:
\begin{equation}\label{a12}
\frac{d}{dx}\frac{\theta}{U^2+\sqrt{1-k^2}}=-\frac{4\lambda_+ U^2}{(U^2+\sqrt{1-k^2})^2},
\end{equation}
which can be integrated to the form
\begin{equation}\label{a12-t}
\theta(x,t) = \left[ U(x)^2+\sqrt{1-k^2} \right] \left[ -4\lambda_+ \int_0^x \frac{U(y)^2}{(U(y)^2+\sqrt{1-k^2})^2} dy + \theta_0(t) \right],
\end{equation}
where $\theta_0$ is a constant of integration in $x$ that may depend on $t$.

For the $cn$-periodic waves with $U(x) = k {\rm cn}(x;k)$, it follows from (\ref{lambda-1-cn}) and (\ref{24-t}) that
\begin{equation}
\label{g2}
2k \left[ P_1(x)^2 - Q_1(x)^2 \right] = 2 U'(x)
\end{equation}
and
\begin{equation}
\label{g3}
2k \left[ P_1(x)^2+Q_1(x)^2 \right]  = 2\lambda_I U(x),
\end{equation}
where $\lambda = \lambda_I$ is used. Together with (\ref{28-t}), we rewrite the differential relation (\ref{50-t}) in the explicit form:
\begin{equation}\label{54}
\frac{d}{dx}\frac{\theta}{|U|^2+ik\sqrt{1-k^2}}=-\frac{4\lambda_I U^2}{(U^2 + ik\sqrt{1-k^2})^2},
\end{equation}
which can be integrated to the form
\begin{equation}\label{54-t}
\theta(x,t) = \left[ U(x)^2 + i k \sqrt{1-k^2} \right]
\left[ -4\lambda_I \int_0^x \frac{U(y)^2}{(U(y)^2 + i k \sqrt{1-k^2})^2} dy + \theta_0(t) \right],
\end{equation}
where $\theta_0$ is a constant of integration in $x$ that may depend on $t$.

We shall now add the time dependence for the function $\theta$. By using
(\ref{periodic-wave-t}) and (\ref{49}), we can write the non-periodic solutions
$(\textsf{p}_1,\textsf{q}_1)$ in the form
\begin{equation}\label{49-t}
\textsf{p}_1(x,t) = \frac{\theta(x,t) - 1}{Q_1(x)} e^{ict/2}, \qquad \textsf{q}_1(x,t) = \frac{\theta(x,t) + 1}{P_1(x)} e^{-ict/2}.
\end{equation}
Substituting (\ref{49-t}) into (\ref{3}) yields the following equation on $\theta$:
\begin{equation*}
\frac{\partial \theta}{\partial t} = i \frac{2 Q_1(x) (U'(x) + 2 \lambda_1 U(x))}{P_1(x)}.
\end{equation*}
By using (\ref{24-t}), this equation can be further rewritten in the form
\begin{equation}\label{50-t-t-t}
\frac{\partial \theta}{\partial t} = 8 i {\rm Re}(\lambda_1) P_1(x) Q_1(x).
\end{equation}

For both $dn$- and $cn$-periodic waves, we substitute either (\ref{a12-t}) or (\ref{54-t}) into (\ref{50-t-t-t})
and use either (\ref{a8-t}) or (\ref{28-t}). Both cases yield
the same equation $\theta_0'(t) = -2i$ with the solution $\theta_0(t) = -2it$,
where the constant of integration in $t$ is neglected due to translational invariance of the NLS equation
(\ref{1}) with respect to $t$.

\subsection{Darboux transformation}

The $N$-fold transformation for the NLS equation was derived and justified in \cite{CP2014}
by using the dressing method. Adopting the present notations with $N = 1$, $\lambda_1 = -iz_1$,
$(p_1,q_1) = \sigma_3 \sigma_1 \bar{s}_1$, where $s_1$ and $z_1$ were used in \cite{CP2014}
and $\sigma_1$ and $\sigma_3$ are standard Pauli matrices, we obtain the one-fold transformation
in the explicit form:
\begin{equation}\label{37}
\tilde{u} = u+\frac{4{\rm Re}(\lambda_1)p_1\bar{q}_1}{|p_1|^2+|q_1|^2}.
\end{equation}
The one-fold transformation (\ref{37}) is fairly well-known for the NLS equation (\ref{1})
(see, e.g., \cite{Sattinger} and references therein). Note that $(p_1,q_1)$ is any nonzero solution
of the linear system (\ref{2})--(\ref{3}) with $\lambda = \lambda_1$.

In order to obtain the two-fold Darboux transformation by using the formalism of \cite{CP2014}, we set
$N = 2$, $\lambda_{1,2} = -iz_{1,2}$, $(p_{1,2},q_{1,2}) = \sigma_3 \sigma_1 \bar{s}_{1,2}$,
and the transformation matrix
$$
M = \left[ \begin{array}{cc} \frac{|p_1|^2 + |q_1|^2}{2 {\rm Re}(\lambda_1)} &
\frac{\bar{p}_1 p_2 + \bar{q}_1 q_2}{\bar{\lambda}_1 + \lambda_2} \\
\frac{p_1 \bar{p}_2 + q_1 \bar{q}_2}{\lambda_1 + \bar{\lambda}_2} &
\frac{|p_2|^2 + |q_2|^2}{2 {\rm Re}(\lambda_2)} \end{array} \right] = \left[ \begin{array}{cc} M_{11} & M_{12} \\
M_{21} & M_{22} \end{array} \right].
$$
By solving the linear system of the dressing method obtained in \cite{CP2014}, we obtain
solutions $r_{1,2}$ of the linear system (\ref{2})--(\ref{3})
with $\lambda_{1,2}$ and the new potential $\tilde{u}$,
where $r_1$, $r_2$, and $\tilde{u}$ are defined in the form:
$$
r_1 = \frac{1}{\det(M)} \left[ \begin{array}{c} \bar{q}_2 M_{12} - \bar{q}_1 M_{22} \\
\bar{p}_1 M_{22} - \bar{p}_2 M_{12} \end{array} \right], \quad
r_2 = \frac{1}{\det(M)} \left[ \begin{array}{c} \bar{q}_1 M_{21} - \bar{q}_2 M_{11} \\
\bar{p}_2 M_{11} - \bar{p}_1 M_{21} \end{array} \right]
$$
and
\begin{equation}\label{40}
\tilde{u} = u + \frac{2 \Sigma}{\det(M)},
\end{equation}
with
\begin{eqnarray*}
\Sigma &=& p_1 \bar{q}_1 M_{22} + p_2 \bar{q}_2 M_{11} - p_1 \bar{q}_2 M_{12} - p_2 \bar{q}_1 M_{21} \\
& = & \frac{p_1 \bar{q}_1 (|p_2|^2+|q_2|^2)}{2 {\rm Re}(\lambda_2)} +
\frac{p_2 \bar{q}_2 (|p_1|^2+|q_1|^2)}{2 {\rm Re}(\lambda_1)} -
\frac{p_2 \bar{q}_1 (p_1 \bar{p}_2 + q_1 \bar{q}_2)}{\lambda_1 + \bar{\lambda}_2} -
\frac{p_1 \bar{q}_2 (\bar{p}_1 p_2 + \bar{q}_1 q_2)}{\bar{\lambda}_1 + \lambda_2}
\end{eqnarray*}
and
\begin{eqnarray*}
\det(M) &=& M_{11} M_{22} - M_{12} M_{21} \\
&=& \frac{(|p_1|^2 + |q_1|^2)(|p_2|^2 + |q_2|^2)}{4 {\rm Re}(\lambda_1) {\rm Re}(\lambda_2)} - \frac{|p_1 \bar{p}_2 + q_1 \bar{q}_2|^2}{|\lambda_1 + \bar{\lambda}_2|^2}.
\end{eqnarray*}
This solution was used in \cite{CP2014} to inspect two-soliton solutions of the NLS equation (\ref{1}).

By using the non-periodic solutions of the linear system (\ref{2})--(\ref{3})
and the Darboux transformations (\ref{37}) and (\ref{40}), we can finally obtain the exact solutions
for the rogue periodic waves of the NLS equation (\ref{1}) in the sense of the definition (\ref{rogue-wave-def}).

\subsection{Rogue $dn$-periodic waves}

Let $u$ be the periodic wave given by (\ref{periodic-wave}) and (\ref{dn-periodic}), while $(p_1,q_1)$ be the $x$-periodic eigenfunction
of the linear system (\ref{2})--(\ref{3}) with $\lambda = \lambda_+$ given by (\ref{lambda-1-dn}).
Substituting (\ref{periodic-wave-t}), (\ref{a8-t}), and (\ref{a6}), into the
one-fold Darboux transformation (\ref{37})
yields a new solution to the NLS equation (\ref{1}) in the form:
\begin{equation}\label{a14}
\tilde{u}(x,t) =-\frac{\sqrt{1-k^2}}{{\rm dn}(x;k)} e^{ict} = -{\rm dn}(x+K(k);k)e^{ict},
\end{equation}
where $K(k)$ is the complete elliptic integral and Table 16.8 in \cite{AS} has been used for
the half-period of the function ${\rm dn}(x;k)$ in $x$. The new solution $\tilde{u}$
is just a translation of the $dn$-periodic wave in $x$, hence it is not a new rogue wave
in the sense of definition (\ref{rogue-wave-def}).

In order to obtain a rogue ${\rm dn}$-periodic wave, we replace $(p_1,q_1)$ in (\ref{37})
by the non-periodic solution $(\textsf{p}_1,\textsf{q}_1)$ of the linear system (\ref{2})--(\ref{3})
with $\lambda = \lambda_+$ given by (\ref{lambda-1-dn}).
Substituting (\ref{a8-t}), (\ref{a6}), (\ref{g1}), and (\ref{49-t}) into the one-fold
Darboux transformation (\ref{37}) yields a new solution to the NLS equation (\ref{1}) in the form:
\begin{eqnarray*}
\tilde{u}(x,t) & = & e^{ict} \left[ U(x) -
\frac{4\lambda_+ (1-2i {\rm Im} \theta(x,t) - |\theta(x,t)|^2) P_1(x) \bar{Q}_1(x)}{
(|\theta(x,t)|^2+1)(|P_1(x)|^2+|Q_1(x)|^2)+2 {\rm Re}\theta(x,t) (|Q_1(x)|^2-|P_1(x)|^2)} \right]\\
& = & e^{ict} \left[ {\rm dn}(x;k) +
\frac{(1-2i {\rm Im} \theta(x,t) - |\theta(x,t)|^2)({\rm dn}(x;k)^2+\sqrt{1-k^2})}{(|\theta(x,t)|^2+1)
{\rm dn}(x;k) + 2 (1 - \sqrt{1-k^2}) {\rm Re}\theta(x,t) {\rm sn}(x;k) {\rm cn}(x;k)} \right],
\end{eqnarray*}
where
\begin{equation}\label{theta-dn}
\theta(x,t) = \left[ U(x)^2+\sqrt{1-k^2} \right] \left[ -4\lambda_+ \int_0^x \frac{U(y)^2}{(U(y)^2+\sqrt{1-k^2})^2} dy -2it \right].
\end{equation}

The new solution $\tilde{u}$ is no longer periodic in $x$. Thanks to the separation of real and imaginary
parts in (\ref{theta-dn}), $|\theta(x,t)| \to \infty$ as $|x| + |t| \to \infty$
everywhere on the plane $(x,t)$, so that
$$
|\tilde{u}(x,t)| \to {\rm dn}(x+K(k);k) \quad \mbox{\rm as} \quad |x| + |t| \to \infty.
$$
Hence, $\tilde{u}$ is a rogue $dn$-periodic wave in the sense of the definition (\ref{rogue-wave-def}).
Similarly to the computations in \cite{ChenPel} one can show that
the maximum of $|\tilde{u}(x,t)|$ occurs at $(x,t) = (0,0)$,
for which we use $\theta(0,0) = 0$ and obtain $|\tilde{u}(0,0)| = 2+\sqrt{1-k^2}$.
Since the maximum of ${\rm dn}(x;k)$ is one, the magnification factor
of the rogue $dn$-periodic wave is $M_{\rm dn}(k) = 2 + \sqrt{1-k^2}$.

\begin{figure}[htb]
\begin{center}
\includegraphics[height=5.cm]{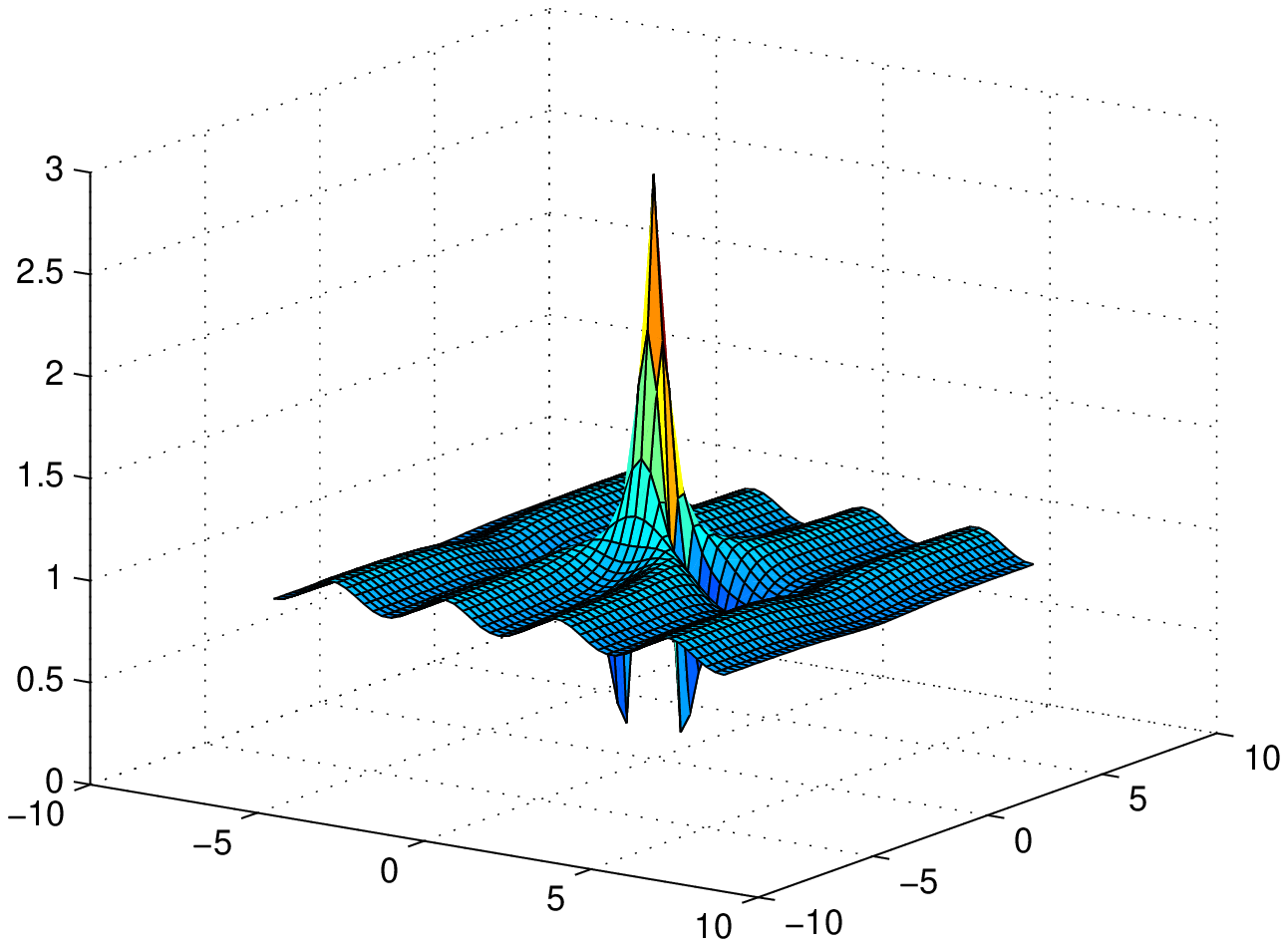}
\includegraphics[height=5.cm]{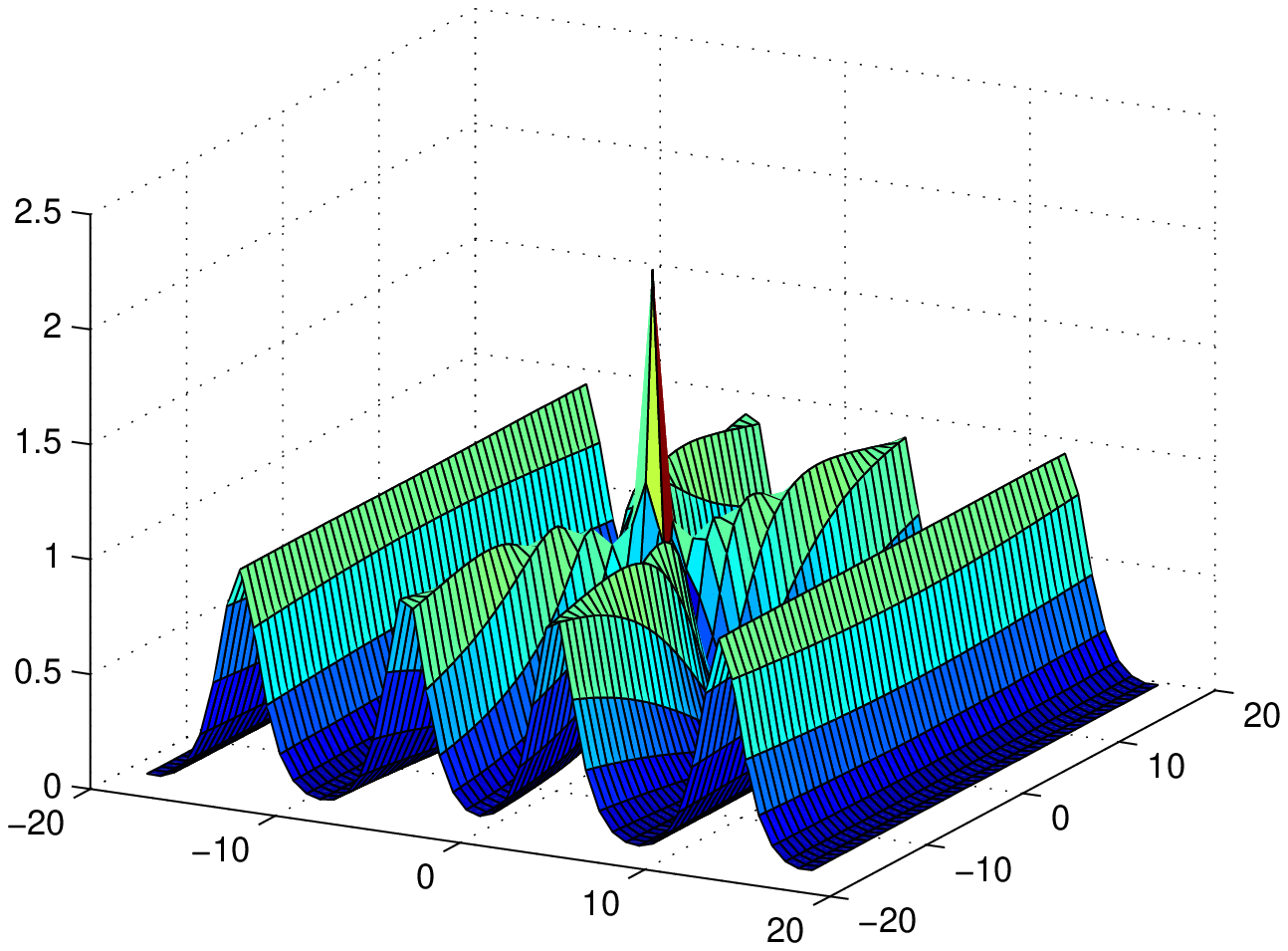}
\end{center}
\caption{The rogue {\em dn}-periodic wave of the NLS for $k = 0.5$ (left) and $k = 0.999$ (right). }
\label{fig-1}
\end{figure}

Figure \ref{fig-1} illustrates
the rogue $dn$-periodic waves for $k = 0.5$ (left) and
$k = 0.999$ (right). In the small-amplitude limit $k \to 0$, the rogue $dn$-periodic wave looks like
the Peregrine's breather (\ref{rogue-basic}) but the wave background is periodic rather than constant.
In the soliton limit $k \to 1$, the rogue $dn$-periodic wave looks like a non-trivial interaction of the two adjacent
NLS solitons (\ref{soliton}). This comparison is confirmed with the limits of the magnification factor
$M_{\rm dn}(k)$. As $k \to 0$, $M_{\rm dn}(k) \to M_0 = 3$, where $M_0$ is the magnification
factor of the Peregrine's breather (\ref{rogue-basic}). As $k \to 1$, $M_{\rm dn}(k) \to 2$
for two nearly-identical NLS solitons (\ref{soliton}) of unit amplitude. The latter result
is in agrement with the recent work \cite{PelSl}, where it was shown in the context of the modified KdV
equation that the magnification factor of the rogue waves built from $N$ nearly identical solitons is
exactly $N$.

Note that the one-fold Darboux transformation (\ref{37}) can be used with the periodic
function $(p_1,q_1)$ defined for $\lambda_1 = \lambda_-$ given by (\ref{lambda-1-dn}).
However, since $U(x)^2 - \sqrt{1-k^2}$ vanishes for some $x \in [0,K(k)]$, the expression
for $\theta$ becomes singular. It is apparently a technical difficulty, which can be resolved,
but we leave this problem for future work.

\subsection{One-fold rogue $cn$-periodic waves}

Let $u$ be the periodic wave given by (\ref{periodic-wave}) and (\ref{cn-periodic}),
while $(p_1,q_1)$ be the $x$-periodic eigenfunction
of the linear system (\ref{2})--(\ref{3}) with $\lambda = \lambda_I$ given by (\ref{lambda-1-cn}).
Substituting (\ref{periodic-wave-t}), (\ref{29}), and (\ref{34}) into the one-fold Darboux transformation (\ref{37})
yields a new solution to the NLS equation (\ref{1}) in the form:
\begin{equation}\label{42}
\tilde{u}(x,t) = -\frac{ik\sqrt{1-k^2} {\rm sn}(x;k)}{{\rm dn}(x;k)} e^{ict} = ik{\rm cn}(x+K(k);k)e^{ict},
\end{equation}
where Table 16.8 in \cite{AS} has been used for the quarter-period of the function ${\rm cn}(x;k)$ in $x$.
The new solution is just a translation of the $cn$-periodic wave $u$
by the gauge and spatial symmetries of the NLS equation (\ref{1}),
hence it is not a rogue wave in the sense of definition (\ref{rogue-wave-def}).

In order to obtain a rogue ${\rm cn}$-periodic wave, we replace $(p_1,q_1)$ in (\ref{37})
by the non-periodic solution $(\textsf{p}_1,\textsf{q}_1)$ of the linear system (\ref{2})--(\ref{3})
with $\lambda = \lambda_I$ given by (\ref{lambda-1-cn}).
Substituting (\ref{29}), (\ref{47}), (\ref{34}), and (\ref{49-t}) into the one-fold
Darboux transformation (\ref{37}) yields a new solution to the NLS equation (\ref{1}) in the form:
\begin{eqnarray*}
\tilde{u}(x,t) & = & e^{ict} \left[ U(x) -
\frac{2k (1-2i {\rm Im} \theta(x,t) - |\theta(x,t)|^2) P_1(x) \bar{Q}_1(x)}{
(|\theta(x,t)|^2+1)(|P_1(x)|^2+|Q_1(x)|^2)+2 {\rm Re}\theta(x,t) (|Q_1(x)|^2-|P_1(x)|^2)} \right]\\
& = & e^{ict} \left[ k {\rm cn}(x;k) + \frac{k(1-2i {\rm Im} \theta(x,t) - |\theta(x,t)|^2)
\left[ {\rm cn}(x;k) {\rm dn}(x;k) + i \sqrt{1-k^2} {\rm sn}(x;k) \right]}{
(|\theta(x,t)|^2+1) {\rm dn}(x;k) + 2 {\rm Re}\theta(x,t) k {\rm sn}(x;k) {\rm cn}(x;k)} \right],
\end{eqnarray*}
where
\begin{equation}\label{theta-cn}
\theta(x,t) = \left[ U(x)^2 + i k \sqrt{1-k^2} \right] \left[ -4\lambda_I \int_0^x \frac{U(y)^2}{(U(y)^2 + i k \sqrt{1-k^2})^2} dy - 2 i t \right].
\end{equation}

The new solution $\tilde{u}$ is no longer periodic in $x$. If
\begin{equation}
\label{tech-constraint}
\int_0^{4K(k)} \frac{U(y)^2 (U(y)^4-k^2(1-k^2))}{(U(y)^2 + k^2 (1-k^2))^2} dy \neq 0
\end{equation}
which is satisfied at least for small $k$, then $|\theta(x,t)| \to \infty$ as $|x| + |t| \to \infty$
everywhere on the plane $(x,t)$, so that
$$
|\tilde{u}(x,t)| \to k |{\rm cn}(x+K(k);k)| \quad \mbox{\rm as} \quad |x| + |t| \to \infty.
$$
Hence, $\tilde{u}$ is a rogue $cn$-periodic wave in the sense of the definition (\ref{rogue-wave-def}).
Similarly to the computations in \cite{ChenPel} one can show that
the maximum of $|\tilde{u}(x,t)|$ occurs at $(x,t) = (0,0)$,
for which we use $\theta(0,0) = 0$ and obtain $|\tilde{u}(0,0)| = 2k$.
Since the maximum of ${\rm cn}(x;k)$ is one, the magnification factor
of the one-fold rogue $cn$-periodic wave is $M_{\rm cn}(k) = 2$ uniformly in $k \in (0,1)$.

\begin{figure}[htb]
\begin{center}
\includegraphics[height=5.cm]{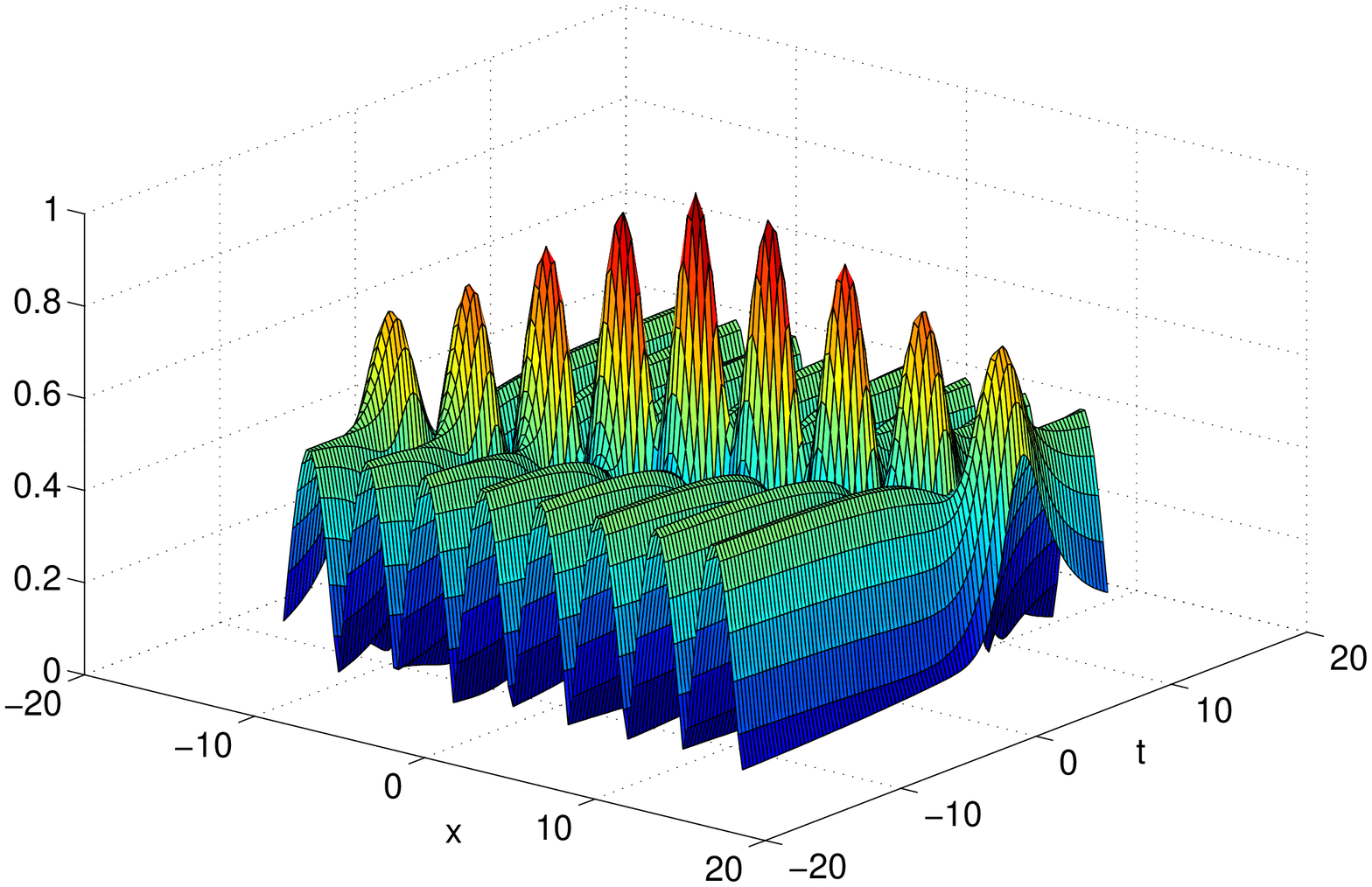}
\includegraphics[height=5.cm]{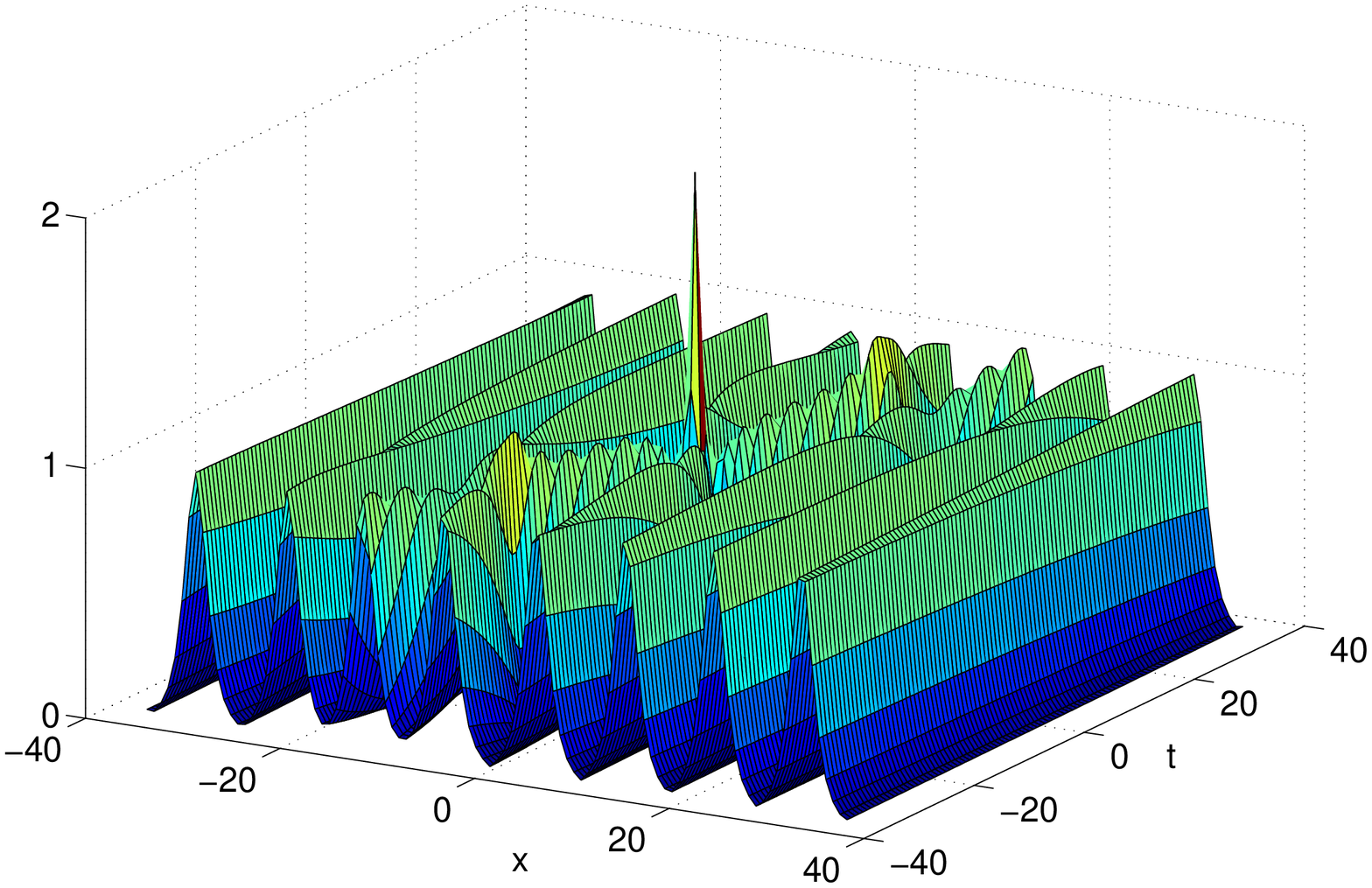}
\end{center}
\caption{The one-fold rogue {\em cn}-periodic wave of the NLS for $k = 0.5$ (left) and $k = 0.999$ (right).}
\label{fig-2}
\end{figure}

Figure \ref{fig-2} illustrates the one-fold rogue $cn$-periodic waves
for $k = 0.5$ (left) and $k = 0.999$ (right).
In the small-amplitude limit $k \to 0$, the rogue $cn$-periodic wave looks like
a propagating solitary wave, however, it is a visual illusion since
the rogue wave is localized in space and time.
In the soliton limit $k \to 1$, the rogue $cn$-periodic wave looks like a non-trivial interaction of the two adjacent
NLS solitons (\ref{soliton}) but it has a different pattern compared to the
interaction of the two adjacent solitons in the rogue $dn$-periodic wave (shown on
the right panel of Figure \ref{fig-1}). It is surprising that
the magnification factor of the one-fold rogue $cn$-periodic wave does not
depend on the amplitude of the $cn$-periodic wave.

The one-fold rogue $cn$-periodic wave does not exist for the modified KdV equation \cite{ChenPel},
because the one-fold Darboux transformation (\ref{37}) with complex $\lambda_1$
produces a complex-valued solution of the modified KdV equation.
In comparison, the NLS equation (\ref{1}) is written for a complex-valued function $u$, hence
the one-fold Darboux transformation (\ref{37}) produces a new rogue $cn$-periodic wave.

\subsection{Two-fold rogue $cn$-periodic waves}

Let us now use the two-fold Darboux transformation (\ref{40}) with $\lambda_2=\bar{\lambda}_1$,
where $\lambda_1 = \lambda_I$ is given by (\ref{lambda-1-cn}).
The periodic eigenfunction $(p_2,q_2)$ is related to the periodic eigenfunction
$(p_1,q_1)$ in (\ref{periodic-wave-t}) by the following relation:
\begin{equation}
\label{periodic-wave-t-second}
p_2(x,t) = \bar{P}_1(x) e^{i c t/2}, \quad q_2(x,t) = \bar{Q}_1(x) e^{-i c t/2},
\end{equation}
Substituting (\ref{28-t}), (\ref{29}), (\ref{47}), and (\ref{34})
into the two-fold Darboux transformation (\ref{40}) yields a new solution to the
the NLS equation (\ref{1}) in the form:
\begin{eqnarray*}
\tilde{u}(x,t) & = & U(x) e^{ict} + \frac{2k \left[ (P_1 \bar{Q}_1 + \bar{P}_1 Q_1) (|P_1|^2 + |Q_1|^2)
- 2 k {\rm Re}[(k+i \sqrt{1-k^2}) P_1 Q_1 (\bar{P}_1^2 + \bar{Q}_1^2)] \right]}
{(|P_1|^2+|Q_1|^2)^2-k^2|P_1^2 + Q_1^2|^2} e^{ict} \\
& = & \left[ k {\rm cn}(x;k) + \frac{2 k {\rm cn}(x;k) \left[ k^2 {\rm cn}(x;k)^2 - {\rm dn}(x;k)^2 \right]}{{\rm dn}(x;k)^2 - k^2 {\rm cn}(x;k)^2} \right] e^{ict} \\
& = & - k {\rm cn}(x;k) e^{ict},
\end{eqnarray*}
which is again a reflection of $u$ by the cubic symmetry.

In order to obtain a rogue ${\rm cn}$-periodic wave, we replace $(p_1,q_1)$ by the non-periodic solution
$(\textsf{p}_1,\textsf{q}_1)$ of the same linear system (\ref{2})--(\ref{3})
with $\lambda_1 = \lambda_I$. The non-periodic solution $(\textsf{p}_1,\textsf{q}_1)$
is given by (\ref{49-t}) with $\theta$ given by the same expression (\ref{theta-cn}).
For $\lambda_2 = \bar{\lambda}_1$, the non-periodic solution $(\textsf{p}_2,\textsf{q}_2)$
is given by
\begin{equation}\label{49-t-second}
\textsf{p}_2(x,t) = \frac{\theta_c(x,t) - 1}{\bar{Q}_1(x)} e^{ict/2}, \qquad
\textsf{q}_2(x,t) = \frac{\theta_c(x,t) + 1}{\bar{P}_1(x)} e^{-ict/2},
\end{equation}
where $\theta_c$ is given by
\begin{equation}\label{theta-cn-second}
\theta_c(x,t) = \left[ U(x)^2 - i k \sqrt{1-k^2} \right]
\left[ -4 \bar{\lambda}_I \int_0^x \frac{U(y)^2}{(U(y)^2 - i k \sqrt{1-k^2})^2} dy - 2 i t \right].
\end{equation}
After some lengthy computations,
we obtain a new solution to the NLS equation (\ref{1}) in the form:
\begin{equation}\label{40-rogue}
\tilde{u}(x,t) = U(x) e^{ict} + \frac{2k N(x,t)}{D(x,t)} e^{ict},
\end{equation}
where
\begin{eqnarray*}
D & = & (|P_1|^2 |\theta - 1|^2+|Q_1|^2 |\theta + 1|^2)(|P_1|^2 |\theta_c - 1|^2+|Q_1|^2 |\theta_c + 1|^2) \\
& \phantom{t} & - k^2 |P_1^2 (\theta - 1) (\bar{\theta}_c - 1) + Q_1^2 (\theta + 1) (\bar{\theta}_c + 1)|^2, \\
N & = & P_1 \bar{Q}_1 (\theta - 1) (\bar{\theta}+1) (|P_1|^2 |\theta_c - 1|^2+|Q_1|^2 |\theta_c + 1|^2) \\
& \phantom{t} & + \bar{P}_1 Q_1 (\theta_c-1)(\bar{\theta}_c + 1)) (|P_1|^2 |\theta - 1|^2+|Q_1|^2 |\theta + 1|^2) \\
& \phantom{t} & - k (k+i \sqrt{1-k^2}) P_1 Q_1 (\theta - 1) (\bar{\theta}_c + 1) (\bar{P}_1^2 (\bar{\theta}-1)(\theta_c - 1)
+ \bar{Q}_1^2 (\bar{\theta}+1)(\theta_c+1)) \\
& \phantom{t} & - k (k-i \sqrt{1-k^2}) \bar{P}_1 \bar{Q}_1 (\theta_c - 1) (\bar{\theta} + 1)
(P_1^2 (\theta - 1)(\bar{\theta}_c-1) + Q_1^2 (\theta+1)(\bar{\theta}_c+1)).
\end{eqnarray*}

Under the same constraint (\ref{tech-constraint}),
$|\theta(x,t)|,|\theta_c(x,t)| \to \infty$ as $|x| + |t| \to \infty$ everywhere on the plane $(x,t)$, so that
$$
\tilde{u}(x,t) \to -u(x,t) \quad \mbox{\rm as} \quad |x| + |t| \to \infty.
$$
Hence $\tilde{u}$ is a new rogue $cn$-periodic wave in the sense of the definition (\ref{rogue-wave-def}).
By using $\theta(0,0) = \theta_c(0,0) = 0$, we obtain $\tilde{u}(0,0) = 3k$.
Since the maximum of ${\rm cn}(x;k)$ is one, the magnification factor
of the two-fold rogue $cn$-periodic wave is $M_{\rm cn}(k) = 3$ uniformly in $k \in (0,1)$.

\begin{figure}[htb]
\begin{center}
\includegraphics[width=7.cm,height=4.cm]{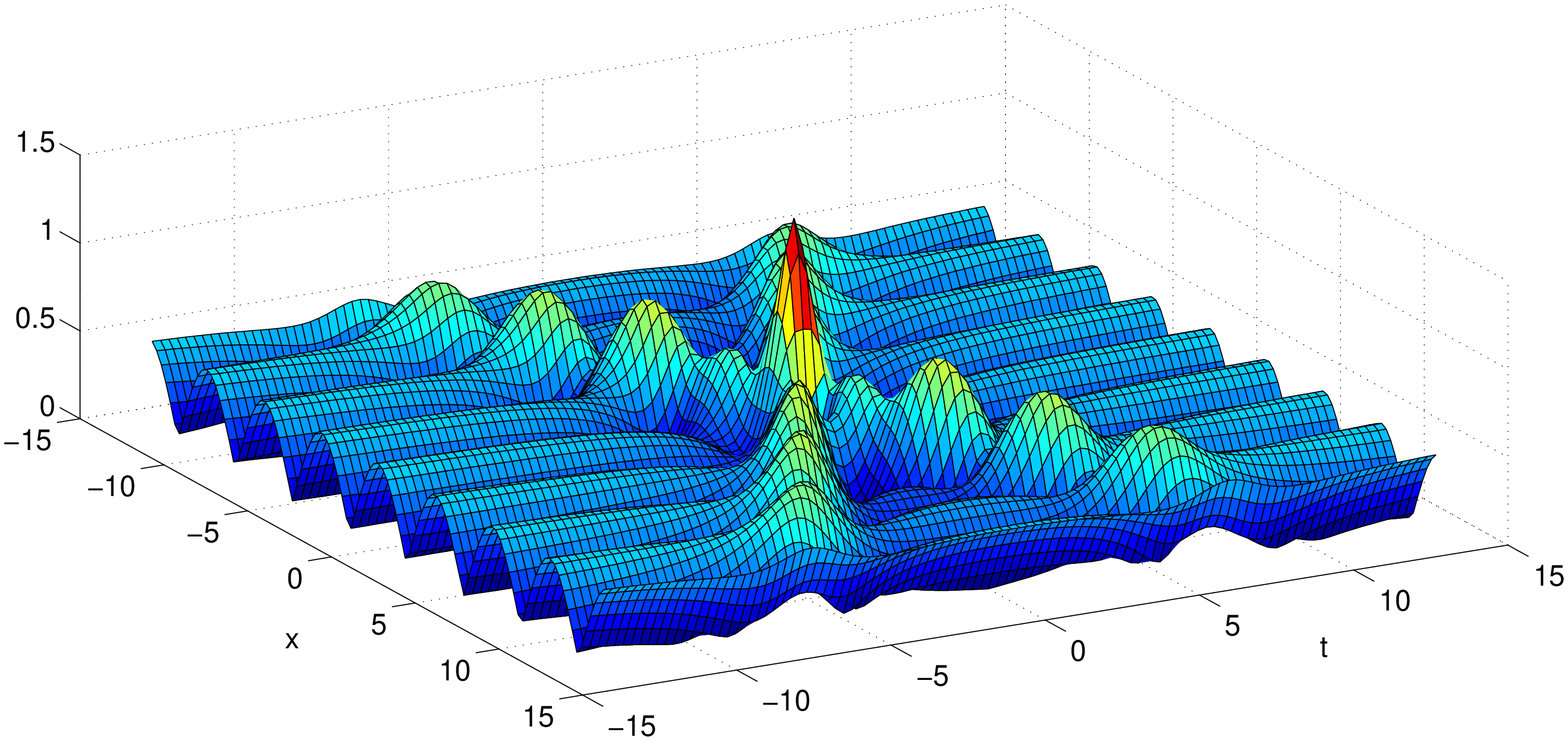}
\includegraphics[width=7.cm,height=4.cm]{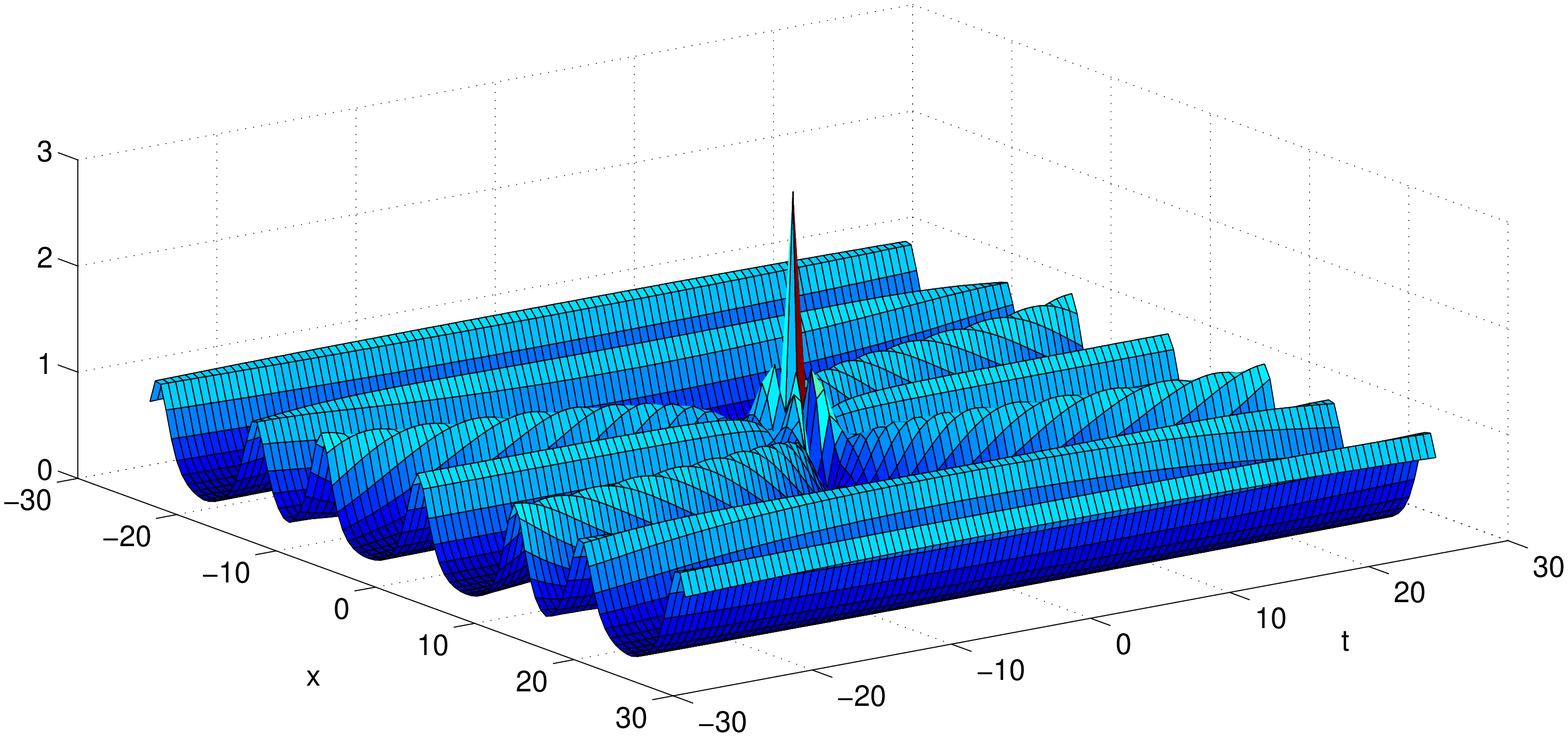}
\end{center}
\caption{The two-fold rogue {\em cn}-periodic wave of the NLS for $k = 0.5$ (left) and $k = 0.999$ (right).}
\label{fig-3}
\end{figure}

Figure \ref{fig-3} illustrates the two-fold rogue $cn$-periodic waves
for $k = 0.5$ (left) and $k = 0.999$ (right).
In the small-amplitude limit $k \to 0$, the rogue $cn$-periodic wave looks like
two propagating solitary waves but they are again localized in space and time.
In the soliton limit $k \to 1$, the rogue $cn$-periodic wave looks like a non-trivial interaction
of the three adjacent NLS solitons (\ref{soliton}) and these explain
why the magnification factor is three, in agreement with the recent work \cite{PelSl}.
It is still surprising that the magnification factor of the two-fold rogue
$cn$-periodic wave does not depend on the amplitude of the $cn$-periodic wave.

\section{Further discussion}

We have developed a computational algorithm of constructing rogue periodic waves
in the context of the focusing NLS equation. Since both $dn$- and $cn$-periodic waves are modulationally unstable,
both waves exhibit rogue waves on their background which appears from nowhere
and disappears without any trace. For the rogue $dn$-periodic waves, we were only able to use one-fold
Darboux transformation since the non-periodic solutions were obtained
in the closed analytical form for only one branch point of the Zakharov--Shabat spectral problem.
For the rogue $cn$-periodic waves, we were able to use both one-fold and two-fold Darboux transformations
because the two branch points in the Zakharov--Shabat spectral problem are related to each other
by complex conjugation and reflection symmetries.

These results can be developed further in view of high interest to rogue waves in the focusing
NLS equation \cite{Tovbis1,Tovbis2,CalSch,Kedziora}. A relatively simple extension of these
solutions would include traveling periodic waves with a nontrivial dependence of the wave phase.
A more difficult problem is to extend the computational algorithm of constructing the rogue waves
for Riemann's Theta functions, which represent quasi-periodic solutions  including
the two-phase solutions considered in \cite{Tovbis1,Tovbis2,CalSch}. These open
questions are left for further studies.

\vspace{0.1cm}

{\bf Acknowledgements:} Jinbing Chen is grateful to the Department of Mathematics
of McMaster University for the generous hospitality during his visit.
The work of J.C. was supported by the National Natural Science Foundation
of China (No.11471072), and the Jiangsu  Overseas Research $\&$ Training Programme
for University Prominent Young $\&$ Middle-aged Teachers and Presidents (No. 1160690028).
The work of D.P. is supported by
the state task of Russian Federation in the sphere of scientific activity (Task No. 5.5176.2017/8.9).

\end{document}